\documentclass[12pt, showpacs,preprintnumbers,amsmath,amssymb]{revtex4}
\usepackage[dvips]{graphicx}
\usepackage{dcolumn}
\usepackage{bm}
\begin{document}
\title{Magnetization of nanomagnet assemblies: Effects of anisotropy and dipolar interactions}
\author{H. Kachkachi}
\email{kachkach@physique.uvsq.fr; http://hamid.kachkachi.free.fr}
\author{M. Azeggagh}
\email{azeggagh@physique.uvsq.fr}
\affiliation{
Laboratoire de Magn\'{e}tisme et d'Optique,
CNRS UMR8634 - Universit\'e de Versailles St. Quentin, \\
45 av. des Etats-Unis, 78035 Versailles, France
}
\date{\today}
\begin{abstract}
We investigate the effect of anisotropy and weak dipolar interactions on the magnetization of an assembly of nanoparticles with distributed magnetic moments, i.e., assembly of magnetic nanoparticles in the one-spin approximation, with textured or random anisotropy.
The magnetization of a free particle is obtained either by a numerical calculation of the partition function or analytically in the low and high field regimes, using perturbation theory and the steepest-descent approximation, respectively. 
The magnetization of an interacting assembly is computed analytically in the range of low and high field, and numerically using the Monte Carlo technique.

Approximate analytical expressions for the assembly magnetization are provided which take account of the dipolar interactions, temperature, magnetic field, and anisotropy.
The effect of anisotropy and dipolar interactions are discussed and the deviations from the Langevin law they entail are investigated, and illustrated for realistic assemblies with the lognormal moment distribution.
\end{abstract}
\maketitle
\section{Introduction}
\label{intro}
In order to investigate the fundamental properties of magnetic nanoparticles and
the novelties they exhibit, new materials had to be made and characterized [see, e.g., the recent review articles \cite{batlab02jpd, skomski03jpc}].
Nowadays, there are mainly two prototypes of nanoparticle samples: 
i) assemblies of, e.g., cobalt, Nickel or maghemite nanoparticles \cite{doretal97acp} embedded in a
non-magnetic matrix with volume distribution and randomly oriented easy
axes, with negligible-to-strong dipole-dipole inter-particle interactions (DDI),
depending on concentration;
ii) isolated single particles of cobalt or nickel measured by the technique
of $\mu$-SQUID \cite{wernsdorfer01acp}.
Technological applications require to some extent ever denser assemblies and thus smaller particles.
However, this leads to a dilemma because small particles become
superparamagnetic at even low temperatures, and an optimum material [with
appropriate anisotropy and other physical parameters] has still to be devised.
Moreover, high density entails strong DDI among the particles, and in technological applications such as magnetic recording, this is an issue of special importance because DDI have been widely recognized as being responsible for the deterioration of the signal-to-noise ratio [see e.g., Refs.~\onlinecite{sharrock90ieee, johnson91jap} and references therein].

Experimentally, investigation of the effect of dipolar interactions in nanoparticle assemblies has revealed many new phenomena pertaining to the
collective behavior of the particles, notably the so-called {\it spin-glass state} at low temperature in concentrated assemblies \cite{doretal99jmmm},\cite{jonetal01prb},\cite{troncetal00jmmm}, owing to the long-range of inter-particle DDI.
It has also been observed that the field behavior of the temperature
$T_{\max}$ at the maximum of the zero-field-cooled magnetization strongly depends on the
concentration of the assembly~\cite{sapetal97prb},\cite{kacetal00jpcm},\cite{ezzirthesis98}.
More precisely, the maximum of $T_{\max}$ as a function of the applied field
observed in dilute samples disappears when the concentration of the
latter is increased.
Today, there arises a more fundamental issue about assemblies of nanoparticles that concerns
the understanding of the interplay between the {\it intrinsic} properties,
such as those pertaining to surface effects, and {\it extrinsic} or collective
effects stemming from the long-range DDI.
Many research groups have experimentally studied this interplay in cobalt and
maghemite particle assemblies.
Measurements of the magnetization at high fields performed on the $\gamma
$-Fe$_{2}$O$_{3}$ nanoparticles \cite{troncetal00jmmm},\cite{ezzirthesis98}, [see \cite{cheetal95prb} for cobalt particles] have shown that the magnetization is strongly influenced by surface effects, depending on the particle size.
For instance, Fig.~1 of Ref.~\cite{kacetal00epjb} shows that 
i) there is a sudden increase of the magnetization as a function of the applied field when
the temperature reaches $70$ K, and the magnetization does not saturate at the
highest available field, i.e., 5.5 T.
ii) there is an important increase of the magnetization at low temperature,
iii) the thermal behavior of the magnetization at 5.5 T is such that the smaller
is the mean diameter of the particle the faster is the increase of the
magnetization at very low temperature.

From the theoretical point of view, the situation involving both surface effects
and dipolar interactions has never been considered so far mainly because of its
tremendous complexities, and also because one has first to understand these two
effects separately.
Needless to say that, already at the static level, no exact analytical treatment
of any kind is ever possible even in the one-spin approximation, i.e., ignoring the
internal structure of the particles and thereby surface effects.
Only numerical approaches such as the Monte Carlo technique can relieve some of this frustration.
Indeed, applications of this technique to the case of Ising dipoles can be found in Ref~\cite{krebin79zpb}.
The same technique has been used in Ref.~\cite{kectro98prb} to study hysteretic properties of monodisperse assemblies of nanoparticles with the more realistic Heisenberg spin model, in the one-spin approximation where each particle carries a net magnetic moment.
In Ref.~\onlinecite{jongar01prb}, the Landau-Lifshitz thermodynamic perturbation theory~\cite{lanlif80} is used to tackle the case of weakly dipolar-interacting monodisperse assemblies of magnetic moments with uniformly or randomly distributed anisotropy axes.
The authors studied the influence of DDI on the susceptibility and specific heat of the assembly.

In the present work, we use the same approach as in Ref.~\onlinecite{jongar01prb} with the objective to study the effect of anisotropy and (weak) dipolar interactions on the field and temperature behavior of the magnetization of a monodisperse and polydisperse assembly of magnetic moments.
For this purpose, we consider an assembly of magnetic moments whose magnitudes are distributed according to a Gaussian or (the more often observed) lognormal law.
The anisotropy is taken as uniaxial and either textured along some reference axis or randomly distributed.
The statistical average of the assembly magnetization is obtained, for weak
DDI, using the thermodynamic perturbation theory, as in Ref.~\cite{jongar01prb}, but here the magnetic field is explicitly included in the assembly magnetic energy.
The low field regime, dealt with in Ref.~\onlinecite{raiste02prb}, is generalized here so as to take account of polydispersity and DDI.
In high fields, the magnetization as a function of temperature and field is computed using the steepest-descent approximation.
In the general range of temperature, field, and anisotropy, the magnetization of a non-interacting assembly is computed exactly by numerical integration of the single-moment (free) partition function or using the Monte-Carlo simulation technique.
For interacting assemblies, we use the Monte Carlo technique.

One of the objectives is to provide ready-to-use (semi) analytical formulae for the
field and temperature dependence of the assembly magnetization that take into account moment and easy axes distributions, and weak dipolar interactions.
Moreover, we investigate the effect of anisotropy and DDI and discuss the validity of the Langevin law, for both textured and random-anisotropy, which is invariably used in the literature to interpret the magnetization measurements on nanoparticle assemblies.
The present work is also an extension of the study in Ref.~\cite{kacgar01epjb} where anisotropy was ignored.

The layout of this paper is as follows: Section \ref{energy} defines the energy and notation. The first new results of the present work appear in  section \ref{magnetization} where we use perturbation theory to derive an analytical expression for the magnetization taking account of DDI. Then, we give approximate expressions in the limiting cases of low and high field regimes, for both a free and interacting particle, and for both monodisperse and polydisperse assemblies. We compare these expressions with the exact numerical calculation, and also discuss the effect of anisotropy.
This section ends with a Monte Carlo calculation and discussion of the corresponding results of more realistic assemblies of magnetic nanoparticles, namely assemblies with the lognormal distribution for the magnetic moments derived from experimental data of nanoparticle assemblies with mean diameter $D_m = 3$ and $7$ nm.
The last section is a conclusion and statement of a few open problems to be dealt with in future investigations.
\section{\label{energy}Energy}
Consider an assembly of magnetic moments ${\bf m}_i = m_i{\bf s}_i,\, i=1,\ldots,{\cal N}$ of magnitude $m_i$ and direction ${\bf s}_i$, with $\vert{\bf s}_i\vert=1$. 
The magnitude of the magnetic moment ${\bf m}_i$ is then defined in terms of the Bohr magneton $\mu_B$, i.e., $m_i=n_i\mu_B$, and the numbers $n_i$ are either all equal for monodisperse assemblies or chosen according to some distribution, the so-called polydisperse assemblies.
Each magnetic moment will have a uniaxial easy axis ${\bf e}_i$, and for an assembly these may be either all directed along some reference axis leading to a textured assembly, or randomly distributed. The latter case will be referred to as random anisotropy.
Hence, the energy of a magnetic moment ${\bf m}_i$ interacting with all the others via DDI, in the magnetic field ${\bf H}=H{\bf e}_h$, reads [after multiplying by $-\beta=-1/k_BT$],  
\begin{equation}\label{osp_energy_1}
{\cal E}_i = \frac{KV_i}{2k_BT}\left({\bf s}_i\cdot {\bf e}_i\right)^{2} + \frac{n_i H\mu_B}{k_BT}\left( {\bf s}_i\cdot {\bf e}_h\right) + \frac{\mu_0 \mu_B^2 n_i}{4 \pi a^{3}k_{B}T}\sum _{j<i}n_j\frac{3({\bf
s}_{i}\cdot{\bf e}_{ij})({\bf s}_{j}\cdot{\bf e}_{ij})-{\bf s}_{i}\cdot{\bf
s}_{j}}{r_{ij}^{3}}, 
\end{equation}
where
\begin{equation}\label{Ed_rij}
{\bf r}_{ij}={\bf r}_{i}-{\bf r}_{j},\, \, \, {\bf e}_{ij}={\bf r}_{ij}/r_{ij}
\end{equation}
is the vector joining the sites $i$ and $j$ and whose magnitude is measured in units of $a$,
a characteristic length on the lattice to be evaluated later on.
Since we are considering assemblies with moment instead of volume distribution, the volume $V_i$ in (\ref{osp_energy_1}) is rewritten in terms of $n_i$ via the saturation magnetization of the material per unit volume $M_s$, i.e., $V_i=m_i/M_s=(\mu_B/M_s)n_i$.
For convenience we introduce the dimensionless parameters
\begin{equation}\label{dimless_params}
x = \frac{\mu_BH}{k_BT}, \quad \sigma = \frac{\mu_BK}{M_s k_BT}, \quad \xi_d = \frac{\mu_0 \mu_B^2}{4 \pi a^{3}k_{B}T},
\end{equation}
and define $x_i=xn_i, \sigma_i=\sigma n_i$. Note that $\sigma_i=KV_i/(k_BT)$ is the commonly used notation for the reduced anisotropy-barrier height of the particle $i$.
Therefore, we rewrite (\ref{osp_energy_1}) as
\begin{eqnarray}\label{Ei_dimless}
{\cal E}_i & = & x_i{\bf s}_i\cdot{\bf e}_{h} + \frac{\sigma_i}{2}({\bf s}_{i}\cdot{\bf e}_{i})^{2} + \xi_d\sum _{j<i}n_i n_j\frac{3({\bf s}_{i}\cdot{\bf e}_{ij})({\bf s}_{j}\cdot{\bf e}_{ij})-{\bf s}_{i}\cdot{\bf s}_{j}}{r_{ij}^{3}} \\ \nonumber
&\equiv& {\cal E}_i^{(0)} + \xi_d\sum _{j<i}n_i n_j\frac{3({\bf s}_{i}\cdot{\bf e}_{ij})({\bf s}_{j}\cdot{\bf e}_{ij})-{\bf s}_{i}\cdot{\bf s}_{j}}{r_{ij}^{3}},
\end{eqnarray}
where ${\cal E}_i^{(0)}$ is the free particle energy. 
In what follows we also occasionally use the dimensionless magnetic moment vector ${\bf S}_i = n_i {\bf s}_i$, and the DDI term is rewritten as a quadratic form in ${\bf S}_i$ 
\begin{equation}\label{DDIterm}
\sum _{j<i}\frac{3({\bf S}_i\cdot{\bf e}_{ij})({\bf S}_j\cdot{\bf e}_{ij})-{\bf S}_i\cdot{\bf S}_j}{r_{ij}^{3}} = \sum _{j<i}{\bf S}_i\cdot{\cal D}_{ij}\cdot{\bf S}_j\equiv \sum _{j<i}\Phi_{ij},
\end{equation} 
where we have introduced the DDI tensor \cite{jongar01prb}
\begin{equation}\label{Ed_tensor}
{\cal D}_{ij}\equiv \frac{1}{r_{ij}^3}\left(3{\bf e}_{ij}{\bf e}_{ij}-1\right).
\end{equation}
\section{\label{magnetization}Magnetization}
\subsection{General formulation}
\label{general_mag}
In order to calculate the thermal-equilibrium average of any observable ${\cal O}({\bf s}_{1},\ldots ,{\bf s}_{\cal N})$ we have to average over each particle's moment direction ${\bf s}_i$, the direction of its easy axis ${\bf e}_i$, and the magnitude $m_i$ of its moment (or equivalently $n_i$).
The average of ${\cal O}$ with respect to spatial orientations of all spins is
\begin{equation}\label{average_total}
\left\langle {\cal O}\right\rangle =\frac{1}{Z}\int {\cal D}\Omega \, e^{\cal E}\, {\cal O}.
\end{equation}
where $$ Z=\int D\Omega \, e^{\cal E},$$
with ${\cal D}\Omega =\prod _{i}d\Omega _{i}=\prod _{i}d^{2}s_{i}/2\pi$, and ${\cal E}=\sum_i{\cal E}_i$.
In particular, the magnetization component along the field taken, for instance, along ${\bf e}_z$, of a single particle $i$ is given by
\begin{equation}\label{direction_average}
\left\langle m^{z}_{i}\right\rangle =\frac{1}{Z}\int {\cal D}\Omega\,
e^{\cal E}\, m^{z}_{i}=m_{i}\left\langle s_{i}^{z}\right\rangle,
\end{equation}
which is a function of the easy-axis direction ${\bf e}_{i}$,
$n_{i}$ and the parameters $\sigma, x, \xi_d$ (or equivalently $K,H,T,\xi_d$).
Next, we infer the magnetization of the assembly per particle as
\begin{equation}\label{assembly_mag}
\left\langle m^{z}_{as}\right\rangle (\sigma ,x, \xi_d)=\frac{1}{\cal N}\int
\frac{d^{2}e_{i}}{2\pi }\sum_{i=1}^{\cal N}w(n_i)\, \left\langle
m^{z}_{i}\right\rangle (m_{i},{\bf e}_{i},\sigma ,x, \xi_d),
\end{equation}
where $w(n_i)$ is some distribution of the Bohr magneton numbers $n_i$.

Therefore one first has to compute the magnetization of a single particle defined in (\ref{direction_average}). 
Analytically, this can only be done in the case of weak DDI using thermodynamic perturbation theory \cite{lanlif80}. This operates by expanding the Boltzmann distribution ${\cal P}=Z^{-1}\exp({\cal E})$ in powers of the interaction parameter $\xi _{d}$, that is
$$
{\cal P}= {\cal P}_{0}(1+\xi_{d}F_{1}+\frac{1}{2}\xi_{d}^{2}F_{2}+\ldots),
$$
where
\begin{equation}\label{W0}
{\cal P}_{0}=\frac{1}{Z_{0}}e^{{\cal E}^{(0)}}
\equiv\prod_{i=1}^{\cal N}{\cal P}_{0}^i
\end{equation}
is the Boltzmann distribution of the non-interacting (free) ensemble, and 
$$
Z_{0}=\prod_{i=1}^{\cal N}\left( \int \frac{d^{2}s_{i}}{2\pi }\,
e^{{\cal E}_i^{(0)}}\right) =\prod _{i=1}^{\cal N}Z_{0}^{i},
$$
with $Z_{0}^{i}$ being the directional partition function of the $i^{th}$ free particle.
For the system considered here, $F_{1}$ and $F_{2}$ are some quadratic, respectively quartic, functionals of ${\bf s}_{i}$.

Therefore, the calculation of the average of an observable ${\cal O}$ is reduced to the calculation of averages with respect to the distribution ${\cal P}_{0}$ of low powers of the spin variables. Henceforth, the average with respect to ${\cal P}_{0}$ will be denoted by $\left\langle .\right\rangle _{0}$.

Consequently, to second order in the interaction parameter $\xi_d$, the average
of any physical observable ${\cal O}$ reads,
\begin{equation}\label{average_int}
\left\langle {\cal O}\right\rangle \simeq \left\langle {\cal O}\right\rangle _{0}+\xi_d\Lambda ^{(1)}+\frac{1}{2}\xi_d^{2}\Lambda ^{(2)}+O(\xi ^{3})
\end{equation}
with,
\begin{equation}\label{Lambda}
\displaystyle\left\lbrace
\begin{array}{ll} 
\Lambda^{(1)}\equiv \left\langle {\cal O}G_{1}\right\rangle _{0}-\left\langle
{\cal O}\right\rangle _{0}\left\langle G_{1}\right\rangle _{0},\\ \\
\Lambda^{(2)}\equiv \left\langle {\cal O}G_{2}\right\rangle _{0}-\left\langle
{\cal O}\right\rangle _{0}\left\langle G_{2}\right\rangle _{0}-2\left\langle
G_{1}\right\rangle_{0}\Lambda ^{(1)},
\end{array}
\right.  
\end{equation} 
where,
$$ 
G_{1}\equiv\sum _{i>j}\Phi_{ij},\\ \\
$$
$$
G_{2}\equiv\sum _{i>j}\Phi_{ij}^{2}
+
\sum _{i>j}\sum_{k>l}\Phi_{ij}\Phi_{kl}q_{ik:jl}q_{il:jk},
$$
with $q_{ik:jl}$ annihilating terms containing duplicate pairs:
$q_{ik:jl}=\frac{1}{2}(2-\delta _{ik}-\delta _{jl})(1+\delta_{ik})(1+\delta
_{jl})$ \cite{jongar01prb}. 
Note that in contrast with the situation in Ref.~\onlinecite{jongar01prb}, the $1^\mathrm{st}$-order averages $\left\langle {\cal O}\right\rangle _{0}$ do not vanish here due to the presence of the external field.

The magnetization of a single particle interacting with all other particles in the assembly can be written to second order in $\xi_d$ as in Eq.~(\ref{average_int}), for the observable $S^{z}_{i}$, that is the (dimensionless) magnetization in the direction of the field taken along the $z$ axis,
\begin{equation}\label{average_dip}
\left\langle S^{z}_{i}\right\rangle \simeq \left\langle S^{z}_{i}\right\rangle
_{0}+\xi_d\Lambda^{(1)} + \frac{1}{2}\xi_{d}^{2}\Lambda^{(2)},
\end{equation}
where $\Lambda^{(1)}$ and $\Lambda^{(2)}$ are obtained from Eq.~(\ref{Lambda}) by setting ${\cal O}$ to $S^{z}_{i}$.

Now, the calculation of $\Lambda^{(1)}$ and $\Lambda^{(2)}$ involves that of averages of products of $S^{z}_{i}$ whose order ranges from $1$ to $5$. 
Introducing the notation
$$
b_i\equiv \left\langle S_i^z\right\rangle_0,
\quad b_i^\prime=\frac{\partial \left\langle S_i^z\right\rangle_0}{\partial x},
$$
the average of an arbitrary degree of $S^{z}_{i}$ is expressed by the function $b_i$ and its $n^{th}$-order derivatives $b_i^{(n)}$ \cite{izyskr88}. 
Restricting ourselves to $1^{st}$-order in $\xi_d$, and thereby to $3^{rd}$-order averages, we have
\begin{eqnarray}
\left\langle S_i^z S_j^z\right\rangle_0 &=& b_i b_j + b_i^\prime\delta_{ij},\\ \nonumber
\left\langle S_i^z S_j^z S_k^z\right\rangle_0 &=& b_i b_j b_k + b_i^\prime b_j\delta_{ik}+ b_{i}'b_{k}\delta_{ij} + b_ib_j^\prime \delta_{jk} + b_i^{\prime\prime}\delta_{ij}\delta_{jk}.
\end{eqnarray}
Next, noting that i) since the field is applied along the $z$ axis the average of the $x,y$ components vanishes, ii) DDI only involve pairs of distinct indices, say $i,j$ so that $\delta_{ij}=0$, Eq.~(\ref{average_dip}) leads to the following expression for the magnetization of an interacting assembly (to first order in $\xi_d$), 
\begin{equation}\label{mz_final}
\left\langle S^{z}_{i}\right\rangle \simeq \langle S_{i}^{z}\rangle_0 +
\xi_d\sum_{k=1}^{\mathcal{N}} \langle S_{k}^{z}\rangle_0 A_{ki}
\frac{\partial<S_{i}^{z}>_{0}}{\partial x},
\end{equation}
with
$$
A_{kl}=\frac{\left[ 3({\bf e}_{h}\cdot{\bf e}_{kl})^{2}-1\right]}{r_{kl}^{3}} 
= {\bf e}_h\cdot{\cal D}_{kl}\cdot{\bf e}_h.
$$

As was discussed in Ref.~\onlinecite{jongar01prb} and confirmed in section \ref{subsec:ddi_effect} below, for non spherical systems, the corrections to the magnetization are largely dominated by the first order contribution to the DDI.

Upon examining Eq.~(\ref{mz_final}) one sees that the magnetization of an interacting particle is written in terms of the magnetization of the free particle and its derivatives.
In order to render this expression more explicit and thereby more useful from a practical point of view, we have to consider some limiting cases where analytical expressions can be derived.
So, next we compute the free-particle magnetization $\langle S_{k}^{z}\rangle_0$ and its derivative with respect to $x$.
\subsection{Limiting cases for the free-particle magnetization: effect of anisotropy}
\label{limits_mag}
The free-particle (reduced) magnetization is given by
\begin{equation}\label{exact_free_mag}
\left\langle s^{z}_{i}\right\rangle_0 = \frac{1}{Z_{0}}\int \prod
_{k=1}^{\mathcal{N}}\left( \frac{d^{2}s_{k}}{2\pi }\,
e^{{\cal E}_k^{(0)}}\right) \, s^{z}_{i} = \frac{1}{Z^{i}_{0}}\int \frac{d^{2}s_{i}}{2\pi }\,e^{{\cal E}_i^{(0)}}\, s^{z}_{i}.
\end{equation} 
This can be computed exactly by numerical integration upon changing to spherical coordinates. More precisely, using (\ref{Ei_dimless}) without the DDI term we rewrite (\ref{exact_free_mag}) as follows
\begin{equation}\label{average_m0_num2}
\left\langle s^{z}_{i}\right\rangle _{0}=\frac{1}{z_{0}}\int _{0}^{2\pi
}d\varphi \int _{-1}^{1}du\, e^{\frac{\sigma_i}{2}y^{2}(u)+x_i u}\, u,
\end{equation}
where
$$
y(u)=u\,p_i +\sqrt{1-u^{2}}\sqrt{1-p_i^2} \cos \varphi, \quad
z_{0}(m_{i},\psi,\sigma ,x)=\int_{0}^{2\pi }d\varphi \int_{-1}^{1}du\,
e^{\frac{\sigma_i}{2} y^{2}(u) + x_iu},
$$
with $p_i\equiv{\bf e}_h\cdot{\bf e}_i$. Eq.~(\ref{average_m0_num2}) is valid for all values of $\sigma$ and $x$ (or $K,H,T$, and $V_i$).
However, numerical integration is very time consuming, and knowing that we have to do this for all particles and then average over the two distributions, renders this expression of little practical interest.
Instead, one may derive sensible analytical expressions in the relevant limiting cases such as low and high field regimes. 
The low field case is dealt with perturbatively while the high field case is treated using the steepest-descent approximation.

Let us now give approximate analytical expressions for the magnetization of a particle in the assembly in these two limiting cases.
\subsubsection{Low field}
The low-field expansion obtained in \cite{raiste02prb} reads (upon introducing the particle index $i$)
\begin{equation}\label{lowfield_mag}
\left\langle s_i^{z}\right\rangle_0^\mathrm{lf} \simeq \frac{1+2S_{i2}P_{i2}}{3} x_i  
- \frac{7+70(S_{i2}P_{i2})^2 + 40S_{i2}P_{i2}-12S_{i4}P_{i4}}{315}x_i^3,
\end{equation}
where,
\begin{equation}\label{S_l_1}
S_{il}(\sigma_i)\simeq
\displaystyle\left\lbrace
\begin{array}{ll}  
\frac{(l-1)!!}{(2l+1)!!}(\frac{\sigma_i}{2})^{l/2}+\ldots, &\sigma_i \ll 1, \\ \\
1 - \frac{l(l+1)}{4\sigma_i}+\ldots, &\sigma_i \gg 1 ,
\end{array}
\right.  
\end{equation} 
and $P_{il} = P_{il}({\bf e}_h\cdot{\bf e}_i)$ are the Legendre polynomials.
For a textured assembly, all of the angular functions $P_{il}$ turn into unity, while for randomly distributed easy axes ${\bf e}_i$ we have  $P_{i2} = P_{i4}=0, P_{i2}^2 = 1/5$. Hence, the magnetization in low field becomes
\begin{equation}\label{averaged_lf}
\displaystyle\left\langle s_i^z \right\rangle^\mathrm{lf}_0\simeq
\displaystyle\left\lbrace 
\begin{array}{ll}
\displaystyle\frac{1+2S_{i2}}{3}x_i - \frac{7+70S_{i2}^2 + 40S_{i2}-12S_{i4}}{315}x_i^3, &\mathrm{textured} \\ \\
\displaystyle\frac{x_i}{3}-\frac{1+2S_{i2}^2}{45}x_i^3, &\mathrm{random}.
\end{array}
\right.  
\end{equation}
It is obvious from these expressions that the magnetization is larger if the anisotropy is textured along the applied field.
\subsubsection{High field: steepest-descent approximation}
The single-particle partition function $Z^i_{0}$ in Eq.~(\ref{exact_free_mag}) reads
\begin{equation}\label{sda_Z0}
Z^i_{0} \propto\int d{\bf s}_i\delta ({\bf s}^{2}_i-1)e^{{\cal E}_i^{(0)}}=\int
d{\bf s}_i\delta
({\bf s}_i^{2}-1)e^{\frac{\sigma_i}{2}({\bf s}_i\cdot{\bf e}_i)^{2}+{\bf x}_i.{\bf
s}_i }
\end{equation}
Temporarily dropping the particle index $i$ for simplicity, $Z^i_{0}$ can be rewritten
using the Hubbard-Stratonovich transformation \cite{fradkin91} which consists in
introducing an auxiliary (vector) field ${\bf\xi}$ and using the Gaussian integration formula
\begin{equation}
\exp \left[ \frac{\sigma }{2}({\bf s}.{\bf e})^{2}\right] =\sqrt{\frac{
\sigma }{2\pi }}\int_{-\infty }^{+\infty }du\exp (-\frac{\sigma }{2}
u^{2}+\sigma ({\bf s}.{\bf e})u),
\end{equation}
so that Eq.~(\ref{sda_Z0}) can be rewritten as
\begin{equation}
Z=\mathrm{Const.}\int due^{-\frac{\sigma }{2}u^{2}}\int d{\bf s}\delta ({\bf s}
^{2}-1)e^{{\bf s}.{\bf \xi}}=\mathrm{Const.}\int due^{-\beta S(u,{\bf \xi})},
\label{partition}
\end{equation}
where ${\bf \xi}$ is the (auxiliary) effective field acting on each magnetic moment,
\begin{equation} \label{xi}
{\bf \xi}={\bf x}+\sigma u{\bf e}, 
\end{equation}
and is the (vector) sum of the applied magnetic field and the anisotropy field; 
$S(u,{\bf \xi})$ is the effective action given by
\begin{equation}\label{action}
S(u,{\bf \xi}) = \frac{\sigma }{2\beta }u^{2}-\frac{1}{\beta }\ln \left[ \int d{\bf s}\delta ({\bf s}^{2}-1)e^{{\bf s}.{\bf \xi}}\right]
= \mathrm{Const.} + \frac{\sigma}{2\beta}u^{2} - \dfrac{1}{\beta}\ln \left[ \frac{\sinh (\xi )}{\xi }\right].
\end{equation}
The variable $u$ appearing in Eq.~(\ref{xi}) is determined through the minimization of $S(u,{\bf \xi})$, and which thereby reads
\begin{equation}
u_0=\frac{B(\xi_0)}{\xi_0}{\bf \xi}_0\cdot{\bf e},
\label{uzero}
\end{equation}
where $B(x)=\coth(x)-\frac{1}{x}$ is the Langevin function. This is a transcendental equation for $u_{0}$ leading to a transcendental equation for ${\bf \xi}_0$,
\begin{equation}\label{EqForxi0}
{\bf\xi}_0 = {\bf x} + \sigma\frac{B(\xi_0)}{\xi_0}({\bf \xi}_{0}\cdot{\bf e}){\bf e}.  
\end{equation}
Next, we use the steepest-descent (or saddle point) approximation \cite{kleinert95} to
compute the partition function and the action of the particle in a high magnetic field, i.e., $x\gg 1$.
This consists in Taylor expanding the action (\ref{action}) around $u_0$, which then may be rewritten as 
\begin{equation}\label{betaf}
\beta S = \beta S_0 + \dfrac{1}{2}\ln(1-X),\quad X=\left(\frac{\sigma B(\xi_0)}{\xi_0}\right)\left[1-\frac{({\bf \xi}_0\cdot{\bf e})^2}{\xi_0^2}\right],
\end{equation}
which can be further expanded. Since, in terms of ($u_0,{\bf\xi}_{0})$, the free energy and the action are equal, the particle magnetic moments is given by
\begin{equation}\label{momentprem}
\left\langle{\bf s}\right\rangle = \left[\frac{\partial (\beta S)}{\partial{\bf x}}\right]
_{u_{0}}, 
\end{equation}
and along the field direction, i.e., ${\bf e}_z = {\bf e}_h = {\bf x}/x$, we get
\begin{equation}
\left\langle s^z\right\rangle = \left\langle {\bf s}\right\rangle\cdot{\bf e}_h = \frac{B(\xi_0)}{\xi_0}({\bf\xi}_0\cdot{\bf e}_h).
\end{equation}

In order to compute $\left\langle s^z\right\rangle$, we must solve (\ref{EqForxi0}) for $\xi_0$, and this can be done perturbatively assuming that the applied field is large and proceeding by successive expansions in $1/x$ and replacing several times $u_{0}$ and $\vec{\xi}_{0}$ by their expressions (\ref{uzero}) and (\ref{EqForxi0}), respectively. 
Note that $\xi_0$ depends on $B(\xi_0)$ but since the Langevin function rapidly saturates to $1$, in the high-field case we may take $B(\xi_0)\cong B(x)$ or even $B(\xi_0)\cong 1-1/x$.
Consequently, we obtain
$$
\xi_0 = x + \sigma p^2\left[1 + \frac{-1 + \frac{3}{2}\left(1-p^2\right) \sigma}{x}\right].
$$ 
Upon using these expressions and reinstating the moment index $i$, we obtain the approximate expression for the free particle's magnetization as an expansion in $1/x$, 
\begin{equation}\label{SDA}
\left\langle s_i^z\right\rangle^\mathrm{hf}_0 \simeq 1-\frac{1}{x_i}+\frac{\Sigma_i}{x_i^2},
\end{equation}
where,
$$
\Sigma_i =\frac{\sigma_i}{2}\left[ 3p_i^2-1+\sigma p_i^2(p_i^2-1)\right],\quad p_i \equiv {\bf e}_h\cdot{\bf e}_i.
$$ 

Eq.~(\ref{SDA}) is only valid at high fields but at all temperatures, though it yields a
better approximation at low temperatures.
More precisely, it is valid for fields $H$ larger than
$$
H_\mathrm{min} = \frac{k_BT}{\mu_B n_i}x_\mathrm{min}, \quad x_\mathrm{min} = \frac{1+\sqrt{\Delta}}{2}, \quad\Delta = 1-4\Sigma_i.
$$
Expression (\ref{SDA}) is much easier to handle than (\ref{average_m0_num2}), and it yields the same result in its range of validity, i.e., for $x \gtrsim x_\mathrm{min}$.

For a textured and random-anisotropy assembly, we infer from Eq.~(\ref{SDA})
\begin{equation}\label{averaged_sda}
\left\langle s_i^z \right\rangle^\mathrm{hf}_0\simeq
\displaystyle\left\lbrace 
\begin{array}{ll}
\displaystyle 1-\frac{1}{x_i}+\frac{\sigma_i}{x_i^2}, &\mathrm{textured} \\ \\
\displaystyle 1-\frac{1}{x_i}-\frac{\sigma_i^2}{15}\frac{1}{x_i^2}, &\mathrm{random}.
\end{array}
\right.  
\end{equation} 

\begin{figure*}[ht!]
\includegraphics*[width=8cm, angle=-90]{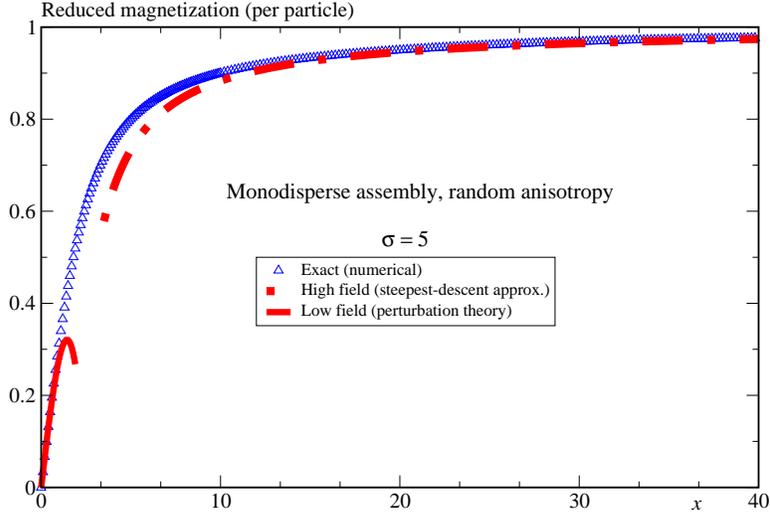}
\caption{\label{asymp_num}
Exact (numerical) calculation versus approximate expressions of the reduced magnetization of a monodisperse assembly with random anisotropy.}
\end{figure*}

The validity of the asymptotic low field and high field expressions of the magnetization given in (\ref{averaged_lf}) and (\ref{averaged_sda}), respectively, is checked by comparing the latter to the exact numerical results obtained from Eq.~(\ref{average_m0_num2}). 
For example, Fig.~\ref{asymp_num} shows such a comparison for a monodisperse assembly and random anisotropy. It is clear that the asymptotic expressions are good enough. Moreover, even at a relatively high temperature (here small $\sigma$), for which the steepest-descent approximation is expected to work worst because $x$ becomes small, expression (\ref{averaged_sda}) renders a good approximation.
%
%
\subsubsection{\label{subsub:comp_langevin}Comparison with the exact calculation and Langevin's function}
We can immediately infer some results from the low and high field expressions derived above. Indeed, considering that the Langevin function ${\cal L}(x_i)=\coth x_i-1/x_i$ expands to $\simeq x_i/3-x_i^3/45$ at low fields and to $\simeq 1-1/x_i$ at high fields, Eqs.~(\ref{averaged_lf}), (\ref{averaged_sda}) readily imply that:
\begin{itemize}
\item both low-field and high-field curves fall onto the Langevin curve if the assembly is isotropic, i.e., $\sigma_i = 0$.
\item the magnetization of a textured assembly falls above the Langevin curve, while that of an assembly with random anisotropy falls below Langevin's curve, and the larger is the anisotropy (hence $\sigma_i$), the larger is the deviation.
\end{itemize}

\begin{figure*}[ht!]
\includegraphics*[width=10cm, angle=-90]{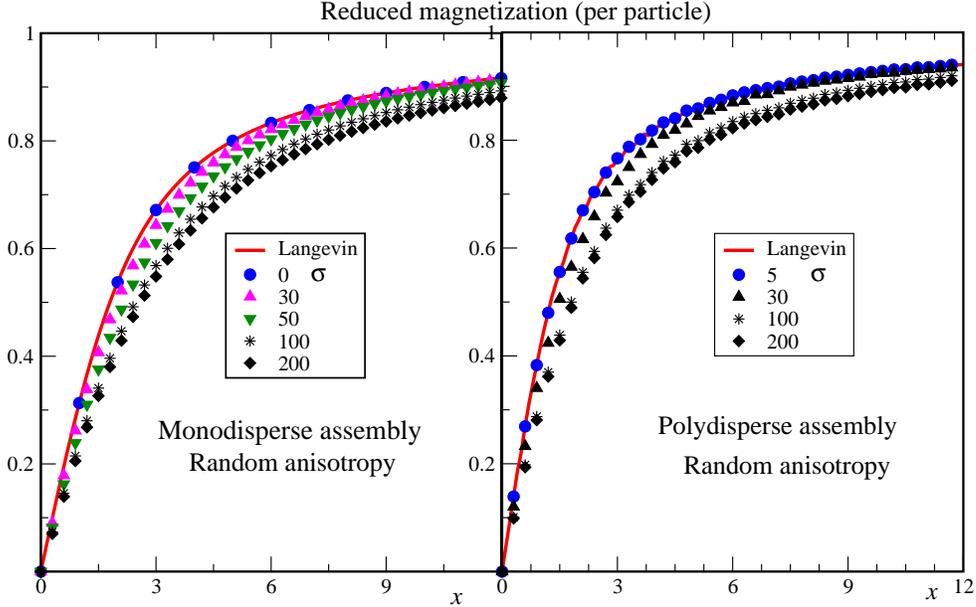}
\caption{\label{osp_ass_unidist_rand}
Reduced magnetization (per particle) of an assembly with randomly distributed anisotropy axes.
Left: for a monodisperse assembly.
Right: for a polydisperse assembly with Gaussian distribution for the magnetic moments.}
\end{figure*}

To confirm these results, we numerically compute the magnetization (\ref{average_m0_num2}) for monodisperse and polydisperse assemblies of magnetic moments, with random anisotropy. In the second case we considered a simple Gaussian (or normal) distribution. The results are presented in Fig.~\ref{osp_ass_unidist_rand}.
%
%
This clearly shows that as $\sigma$ increases, or equivalently at a fixed temperature and increasing anisotropy constant, the magnetization drops. Indeed, stronger anisotropy implies that it is more favorable for the magnetic moments to align along their randomly oriented easy axes, and so the Zeeman energy is not sufficient to align them along the field direction, as is clearly shown by Eq.~(\ref{averaged_lf}), (\ref{averaged_sda}).
This holds for both monodisperse and polydisperse assemblies.
Note also that the zero or very weak anisotropy curves coincide with the Langevin function ${\cal L}(\xi)=\coth(\xi)-1/\xi$, which simply confirms the fact that Langevin's law is only rigorously valid in the absence of anisotropy or at high temperature, or more precisely in the superparamagnetic regime.
On the contrary, for a textured assembly [whose results are not shown here] all easy axes are parallel and obviously stronger anisotropy leads to larger magnetization.

We have also investigated the effect of anisotropy on the magnetization of some typical nanoparticle assemblies with the most often observed volume (or $n$) distribution in samples of magnetic nanoparticles, namely the lognormal distribution with parameters $\mu, \delta$, 
\begin{equation}\label{lognormal}
w(n)=\frac{1}{n\delta\sqrt{2\pi}}\exp \left[-\frac{1}{2}\left(\frac{\ln n-\mu }
{\delta }\right) ^{2}\right].
\end{equation} 
We consider assemblies of, e.g., cobalt or maghemite, nanoparticles with mean diameter $D_m=3$ and $7$ nm.
We simulate the assembly as a collection of magnetic moments randomly assigned to the sites of a regular simple cubic lattice. The moment magnitude distribution is taken from experiments \cite{prenethesis95} upon converting the volume or diameter to the corresponding number of Bohr magnetons $n$.
%
\begin{figure}[!ht]
\includegraphics[width=10cm, angle=-90]{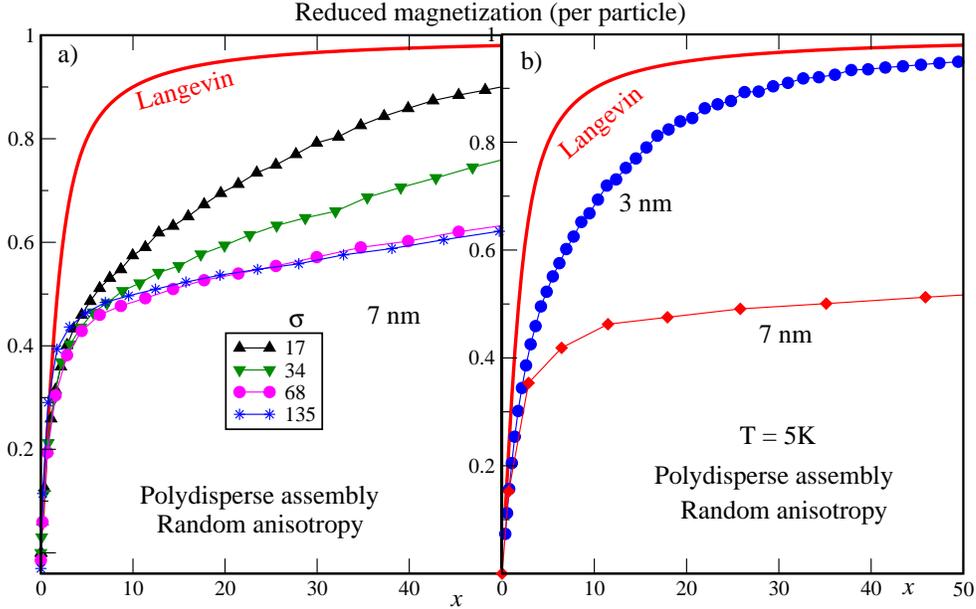}
\caption{\label{poly_ass_size_effect}
a) Reduced magnetization (per particle) of an assembly of ${\cal N}=1024$ with lognormal-distributed magnetic moments with mean diameter $D_m = 7$ nm and randomly-distributed easy axes as obtained from Monte Carlo calculations for different anisotropy values. $x_m=n_m\mu_BH/k_BT$, where $n_m$ is the mean number of Bohr magnetons for this assembly.
b) Langevin function together with the Monte Carlo results for $D_m = 3$ and $7$ nm.}
\end{figure}
%
The magnetization of these assemblies is computed using the standard equilibrium Monte Carlo technique \cite{binher92, kacetal00epjb} at arbitrary temperature, applied field, and DDI parameter $\xi_d$.
For a non-interacting assembly the field behavior of the magnetization, for different values of the parameter $\sigma$ is shown in Figs.~\ref{poly_ass_size_effect}.
In Fig.~\ref{poly_ass_size_effect}a we observe that, in the high field regime, the conclusions drawn from Fig.~\ref{osp_ass_unidist_rand} are confirmed in the present case too. That is, the higher is $\sigma$, the lower is the magnetization. 
On the other hand, in low fields this is not globally so because the competition between Zeeman, thermal and anisotropy contributions to the energy, for this distribution, results in a crossing between the various magnetization curves, as has been observed, e.g., for maghemite particles [see \cite{troncetal00jmmm} (Fig.~10) and \cite{ezzirthesis98}]. 
In fact, this situation is reminiscent of the two phases [blocked and superparamagnetic] exhibited by the zero-field-cooled magnetization, and separated by the temperature $T_\mathrm{max}$ at the peak.
More precisely, at a given applied field $H$, if $x$ is sufficiently large, $T$ is smaller than $T_\mathrm{max}$, and thus the larger is the anisotropy and/or particle volume, the higher is the energy barrier, and the smaller is the magnetization, because most of the particles remain in their blocked states with almost randomly oriented moments.
In the opposite situation, small $x$ corresponds to high temperature, and in this case anisotropy has a negligible effect, so that a higher volume corresponds to a higher magnetic moment (or $n$) and thereby to a higher Zeeman energy, which is necessary to take over the strong thermal fluctuations. Higher Zeeman energy, of course, builds larger magnetization.
Equally important is the observation of the related effect, also observed experimentally in maghemite particles [see e.g., Fig.~10 of Ref.~\onlinecite{troncetal00jmmm}], that the magnetization as a function of field [see Fig.~\ref{poly_ass_size_effect}a] has a much larger slope for small $x$ than for large $x$. 
Furthermore, we see that there is a large deviation from the Langevin law, due to several parameters, ignored by the latter, especially anisotropy.
In addition, we see in Fig.~\ref{poly_ass_size_effect}b that, as was observed earlier, the larger is the mean diameter of the assembly, the larger is $\sigma$, and thereby the larger is the deviation from the Langevin curve.
\subsection{\label{subsec:ddi_effect}Effect of dipolar interactions}
Now, we derive the expressions analogous to Eqs.~(\ref{averaged_lf}), (\ref{averaged_sda}) for a weakly interacting polydisperse assembly. We only do this in the case of random anisotropy.
Accordingly, inserting the low and high field expansions (\ref{averaged_lf}), (\ref{averaged_sda}) in Eq.~(\ref{mz_final}) leads to analytical expressions for the magnetization of a weakly interacting assembly as a function of field, temperature, anisotropy, and the DDI parameter $\xi_d$.
The assembly (reduced) magnetization per particle is defined as $\left\langle s_z \right\rangle_\mathrm{ass}=1/{\cal N}\sum_{i=1}^{\cal N}\left\langle s_z \right\rangle_i$. In the case of randomly distributed easy axes we obtain,
\begin{equation}\label{lfhf_ddi_mag}
\left\langle s_z \right\rangle_\mathrm{ass}\simeq
\displaystyle\left\lbrace 
\begin{array}{ll}
\left[1 + \dfrac{\tilde{\xi}_d}{3}{\cal C}^{(1,2)}\right]\dfrac{\left\langle x\right\rangle}{3}
-\left[A_3 + \dfrac{4}{3}\tilde{\xi}_d A_5 \right]\dfrac{\left\langle x\right\rangle^3}{45} , &\mathrm{low\,field} \\ \\
1-\dfrac{1}{\left\langle x\right\rangle}-\left[\dfrac{\left\langle\sigma\right\rangle^2}{15}-\tilde{\xi}_d{\cal C}^{(0,1)}\right]\dfrac{1}{\left\langle x\right\rangle ^2} + \tilde{\xi}_d\left[\dfrac{2\left\langle\sigma\right\rangle^2}{15}{\cal C}^{(1,1)} - {\cal C}^{(0,0)}\right]\dfrac{1}{\left\langle x\right\rangle ^3}, &\mathrm{high\,field},
\end{array}
\right.  
\end{equation}
where we have defined [see Eq.~(\ref{dimless_params}) for notation] 
$
\tilde{\xi}_d \equiv \xi_d\left\langle n\right\rangle^2, \left\langle x\right\rangle\equiv \left\langle n\right\rangle x, \left\langle \sigma\right\rangle\equiv \left\langle n\right\rangle \sigma$, with $\left\langle n \right\rangle\equiv 1/{\cal N}\sum_{i=1}^{\cal N}n_i,$ 
and the (scaled) constants
\begin{eqnarray}\label{latticeconstants}
{\cal C}^{(a,b)} = \dfrac{1}{\cal N}\sum_{i,j=1, i\neq j}^{\cal N}\dfrac{n_i^aA_{ij}n_j^b}{\left\langle n\right\rangle^{a+b}}, \quad 
A_3 =\dfrac{1}{\cal N}\sum_{i=1}^{\cal N}\dfrac{n_i^3 \alpha_i}{\left\langle n \right\rangle^3}, \quad 
A_5 = \dfrac{1}{\cal N}\sum_{i,j=1, i\neq j}^{\cal N}\dfrac{\alpha_i n_i^3 A_{ij}n_j^2}{\left\langle n \right\rangle^5}.
\end{eqnarray}  
$\alpha_i=1+2S_{i2}^2$ with $S_{i2}$ defined in (\ref{S_l_1}).

Note that there are three types of contributions: There are pure anisotropy terms, pure DDI terms, and mixed terms.
It is readily seen that in the absence of anisotropy and DDI, i.e., for $\sigma_i=0,\xi_d=0$, and from Eq.~(\ref{S_l_1}) $\alpha_i=1$, expressions (\ref{lfhf_ddi_mag}) simplify back, in the case of a monodisperse assembly with $n_i = \bar{n}$, to the expansions of the Langevin function in low and high field regime given at the beginning of section \ref{subsub:comp_langevin}, or as can be inferred from Eqs.~(\ref{averaged_lf}), (\ref{averaged_sda}).
Let us now discuss the constants ${\cal C}^{(a,b)},A_3,A_5$.
%
\begin{figure*}[ht!]
\includegraphics*[width=12cm]{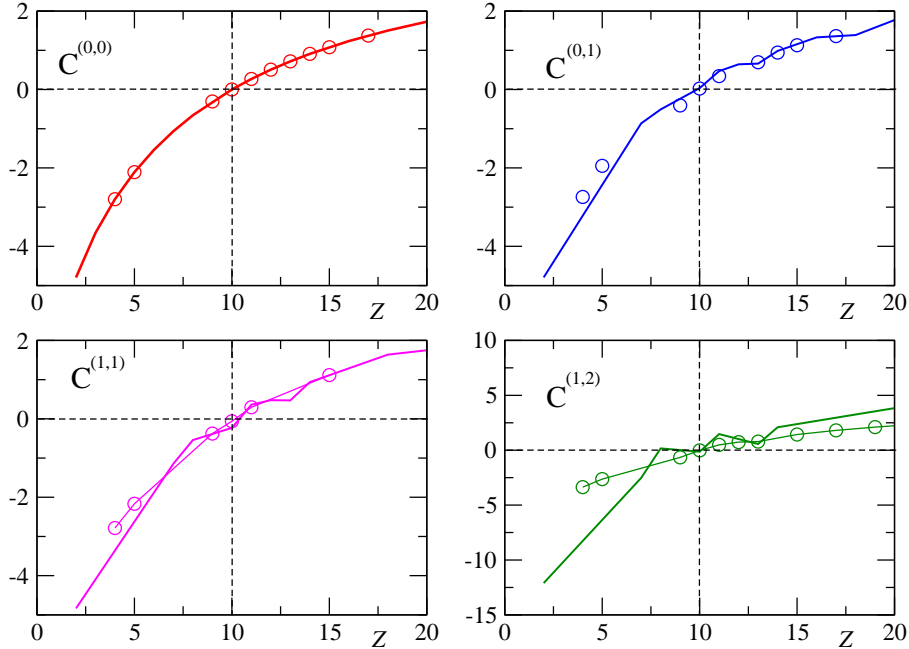}
\caption{\label{Cab_XY10Z}
Some of the constants ${\cal C}^{(a,b)}$ of Eq.~(\ref{latticeconstants}) versus the vertical size $Z$ of the cubic lattice $10\times 10\times Z$. For $Z<10$, the lattice is oblate, for $Z>10$ it is prolate, and for $Z=10$ it is cubic. These results are for a cobalt assembly with $D_m = 3$ nm (thick lines), and $7$ nm (symbols), corresponding to $n_m =(M_sV_m/\mu_B)\simeq 2172$ and $n_m\simeq 27595$, respectively, where we have used $V_m=\pi D_m^3/6$, and $M_s\simeq 1425\times 10^3$ J/T/m$^3$.}
\end{figure*}

%
First, in the continuum limit the lattice sum ${\cal C}^{(0,0)}=\dfrac{1}{\cal N}\sum_{i,j=1, i\neq j}^{\cal N}A_{ij}$ becomes \cite{jongar01prb} ${\cal C}^{(0,0)} = 4\pi (1/3-\lambda _z)$, for a simple cubic lattice, with $\lambda_z$ being the demagnetizing factor along the $z$ axis. For instance, for a box with semi-axes $a=b=5,c=10$, we have
$$
\lambda_{z}=\frac{abc}{2}\int _{0}^{\infty }\frac{ds}{(c^{2}+s)\sqrt{(a^{2}+s)(b^{2}+s)(c^{2}+s)}}\simeq 0.174,
$$
so that ${\cal C}^{(0,0)}=4\pi (1/3-\lambda_z)\simeq 2$.
Note that in the monodisperse case we have ${\cal C}^{(a,b)}\propto {\cal C}^{(0,0)}$.

In the polydisperse case, using the assemblies of Fig.~\ref{poly_ass_size_effect}, in Fig.~\ref{Cab_XY10Z} we plot these constants as functions of the size $Z$ along the field direction for a box-shaped lattice with size ${\cal N}=X\times Y\times Z$, and $X=Y=10$, for the mean diameter of $3$ and $7$ nm.
It is clear from Fig.~\ref{Cab_XY10Z} that these constants, and thereby the corresponding DDI terms in Eq.~(\ref{lfhf_ddi_mag}), are shape dependent. This is no surprise knowing that the long range DDI lead to shape dependence of the physical quantities, and in particular the magnetization.
On the other hand, we see that these constants are negative for the oblate system, and positive for the prolate, which implies that the DDI suppress the assembly magnetization in the former case and enhance it in the latter.
Moreover, it is also seen that for cubic systems all the constants ${\cal C}^{(a,b)}$ vanish, which means that the DDI do not contribute to the magnetization in this case, and thus the deviations from the Langevin behavior are caused only by anisotropy.
Note that all the scaled constants ${\cal C}^{(a,b)}$ are almost independent of the assembly mean diameter. In particular, this is trivial for ${\cal C}^{(0,0)}$.

In order to study the constants $A_3,A_5$ in a similar way, we have to make assumptions about the intensity of the anisotropy, since these constants contain the parameter $\alpha_i$.
From Eq.~(\ref{S_l_1}) we infer that in the absence of anisotropy, i.e., $\sigma_i=0$, $\alpha_i=1$, while for strong anisotropy we may approximate $S_{i2}$ to $1$, and hence $\alpha_i$ to $3$, so that $A_3\propto (1/{\cal N})\sum_i n_i^3/\left\langle n \right\rangle^3$ and $A_5\propto{\cal C}^{(2,3)}$ in both limits of anisotropy. For $D_m=3,7$ nm, $A_3\simeq 2,6$. In the continuum limit $A_3$ tends to $\exp(3\delta^2)$, where $\delta$ is the standard deviation of the distribution (\ref{lognormal}). $A_5$ shows the same behavior as ${\cal C}^{(1,2)}$ but with bigger change with $D_m$. It is well known from other areas of physics that the calculation of such high-order moments (or ``cumulants") requires more precision because they present more statistical fluctuations with the lattice size.
%
%

\begin{figure*}[ht!]
\includegraphics[width=8cm, angle=-90]{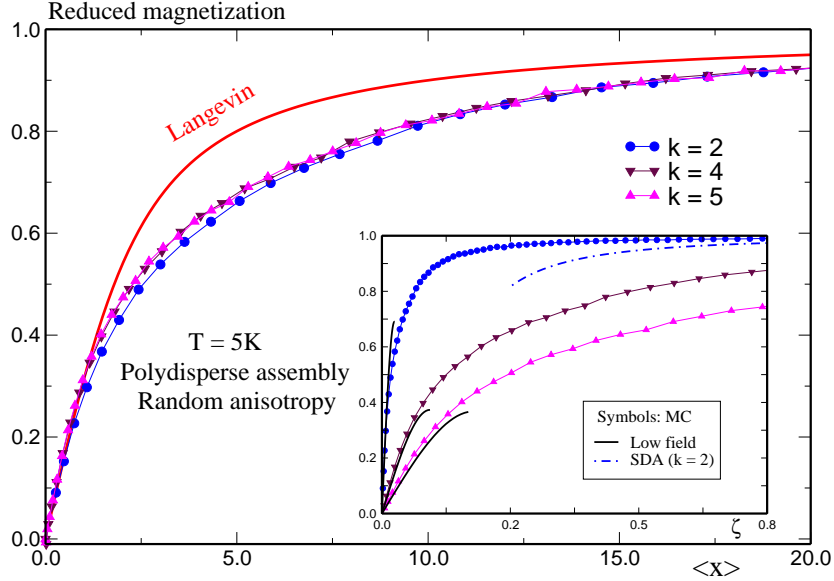}
\caption{\label{osp_ass_ddi_mc_sda}
Reduced magnetization (per particle) of an interacting assembly of ${\cal N} = 10\times 10 \times 5$ lognormal-distributed magnetic moments with mean diameter $Dm = 7 nm$ and random anisotropy.  Monte-Carlo in symbols and analytical expressions (\ref{lfhf_ddi_mag}) in lines. The parameters $\zeta$ and $k$ are defined in the text.}
\end{figure*}

In Fig.~\ref{osp_ass_ddi_mc_sda} we plot the Langevin function (full line) and the Monte Carlo results (symbols) for the magnetization of an interacting assembly of (${\cal N} = 10\times 10 \times 5$) lognormal-distributed moments, with random anisotropy, and for different values of the inter-particle distance.
Here we use the same assemblies as in Fig.~\ref{poly_ass_size_effect}.
The intensity of DDI, or equivalently the value of $\xi_d$, is varied by varying the lattice parameter $a$ entering $\xi_d$ [see Eq.~(\ref{dimless_params})].
More precisely, the parameter $a$ is taken as a real number times the mean diameter $D_m$ of the assembly, i.e., $a=k\times D_m$. Thus, large values of $k$ correspond to an isotropically inflated lattice with large distances between the magnetic moments, and thereby weak DDI.

These results, obtained for an assembly on a simple cubic lattice, do confirm that DDI suppress the magnetization. Indeed, we recall that it was shown by Luttinger and Tisza \cite{luttis46pr} [see also the more recent work \cite{kectro98prb} using the Monte Carlo technique] that the ground state of a simple cubic lattice of dipoles is antiferromagnetic, while that of a face-centered cubic lattice is ferromagnetic.
%
%
It is also seen that these curves deviate from the Langevin law. However, we emphasize that the deviations induced by DDI are much smaller than those induced by anisotropy, as already discussed earlier [see also Ref.~\onlinecite{jongar01prb} for a related discussion of the effect of the system shape on the magnetic susceptibility of a monodisperse assembly].
In the inset of Fig.~\ref{osp_ass_ddi_mc_sda} the same results are magnified by plotting them in function of $\zeta = (\left\langle x \right\rangle/\tilde{\xi}_d)\times 10^{-3}\propto \mu_BH/(\mu_B^2/a^3)$, i.e., the ratio of Zeeman energy to the DDI energy, which also makes it possible to distinctively plot the analytical expressions (\ref{lfhf_ddi_mag}) for low field. In the case of high field only one curve ($k=2$, i.e., relatively strong DDI) is presented since for the other values of DDI parameter $k$, the steepest-descent approximation is valid for much higher values of $\zeta$.
Note also that in function of the parameter $\zeta$ the tendency with increasing DDI strength is reversed.
\section{Concluding remarks and open problems}
\label{conclusion}
We have provided simple approximate analytical expressions for the magnetization of a weakly interacting polydisperse assembly of magnetic moments with randomly distributed easy axes, in both low and high field regimes. These expressions have been checked against extact calculations using either numerical integration of the partition function or Monte Carlo simulations.
We have also computed the magnetization of such systems for an arbitrary inter-particle separation, or equivalently arbitrary intensity of dipole-dipole interactions, using the Monte Carlo technique.
However, in this case only assembly of limited sizes could be dealt with, as the calculation speed is reduced to a crawl by the long-ranged DDI. 
We have investigated the deviations caused by random anisotropy, DDI, and polydispersity, from the Langevin law that is commonly invariably used in the literature to describe the magnetization of real materials.
We have also found that more realistic assemblies with a lognormal volume distribution, render a magnetization that exhibits two different regimes as a function of the applied field, with different variation slopes, as has been observed in experiments on maghemite particles.
Moreover, as a byproduct, we find that the magnetization of an assembly of nanoparticles in the one-spin approximation considered here, i.e., with each particle represented by a macro-spin, does saturate in high fields. This suggests that the magnetization non-saturation observed in experiments on small particles is most likely due to some intrinsic properties of the particles, such as surface effects, as has been argued in many publications [see e.g., \cite{troncetal00jmmm, kacdim02prb} and many references therein].

For future investigation, we intend to apply the Fast-Fourier-Transform technique to speed up the Monte Carlo calculations especially for interacting assemblies of more reasonable sizes, and take account, \textit{inter alia}, of random spatial distributions of the particles on the lattice.
Using the kinetic Monte Carlo technique, we also intend to investigate the disappearance of the maximum in $T_\mathrm{max}(H)$ as the concentration of the assembly is increased, and also the appearance of the spin-glass like state at low temperature. For the latter purpose, we will most likely have to tackle the problem of an assembly of multi-spin particles.
\acknowledgements
We thank Yu. Raikher for reading the manuscript and suggesting judicious improvements.
%

\end{document}